# Unusual role of ligand states in the electronic properties of a parent Fe-based superconductor, $CaFe_2As_2$

Ram Prakash Pandeya, Arindam Pramanik, A. Thamizhavel and Kalobaran Maiti[a]

Department of Condensed Matter Physics and Material Science, Tata Institute of Fundamental Research, Homi Bhabha Road, Colaba, Mumbai – 400005, India.

[a] Corresponding author: kbmaiti@tifr.res.in

**Abstract.** We investigate the role of ligand states in the electronic properties of $CaFe_2As_2$ using high-resolution hard *x*-ray photoemission spectroscopy (HAXPES) at different sample temperatures. Experimental results indicate that the binding energy of Ca is close to that for 2+ charge state of Ca atoms and the other constituent elements, Fe and As possess electronic configuration close to that in elemental systems. No difference is observed in the As 3*p* core level spectra with the change of emission angle and/or the change in sample temperature. This is surprising as the Ca atoms at the cleaved sample surface reorganizes itself to form linear structures which is expected to influence Ca-As hybridization leading to significant difference in surface and bulk electronic structures. Moreover, $CaFe_2As_2$ undergoes structural and magnetic phase transition at 170 K, and strong Fe-As hybridization provides pathways for electron dynamics. Clearly, further studies are required to resolve these puzzling observations.

## INTRODUCTION

Fe-based superconductors have drawn much attention due to their complex electronic properties and the discovery of superconductivity in materials containing magnetic element, Fe, which is unusual. $CaFe_2As_2$ is one such material belonging to '122' class and is one of the most studied parent compounds of Fe-based systems exhibiting rich temperature-pressure and temperature-carrier concentration phase diagram. It forms in paramagnetic tetragonal phase at room temperature and becomes orthorhombic below 170 K along with spin density wave (SDW) magnetic ground state [1]. Application of even a small pressure (~ 0.4 GPa) leads to superconductivity along with a structural transition to a phase having tetragonal symmetry, which is known as collapsed tetragonal phase (cT). In this structural phase, no magnetic long-range order was observed [2]. There are significant debates on the feasibility of superconductivity under pressure; while some studies support such behavior, some do not find such behavior [3]. It has been observed that sample preparation condition influences the electronic and magnetic properties of this system significantly [4]. Detailed electronic structure study using various forms of photoelectron spectroscopy such as high-resolution photoemission spectroscopy (HRPES) and angle-resolved photoemission spectroscopy (ARPES) revealed evidence of cT phase even in samples under ambient condition presumably due to strain [5-7].

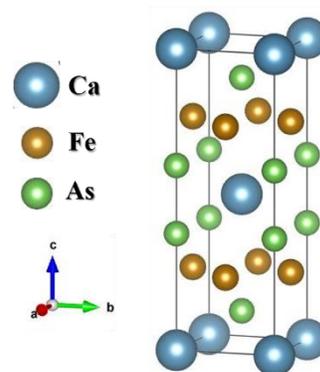

Fig. 1 Room temperature crystal structure of $CaFe_2As_2$

Room temperature crystal structure of $CaFe_2As_2$ is shown in Fig. 1, which is a layered structure. Fe-layers are sandwiched by As layers and the electron conduction happens via hybridization of Fe 3*d*-As 4*p* states. FeAs layers are separated by insulating Ca layers as shown in the figure. Band structure calculations [6] show signature of finite hybridization of Ca 4*s* states with As 4*p* states. Valence band is primarily constituted by the Fe 3*d* states with significant hybridization with As 4*p* states. As 4*p* states provide pathways for electron conduction and the influence of Ca layer on the electronic properties occur via Ca-As hybridization. Clearly, hybridization with As 4$p$ states (ligand states) is the key to derive electronic properties of this system. To study the role of these ligands in the electronic properties in this system, we have employed hard $x$-ray photoemission spectroscopy (HAXPES) and studied the evolution of ligand states as a function of temperature and surface sensitivity of the technique. The results unveil unusual scenario as given below.

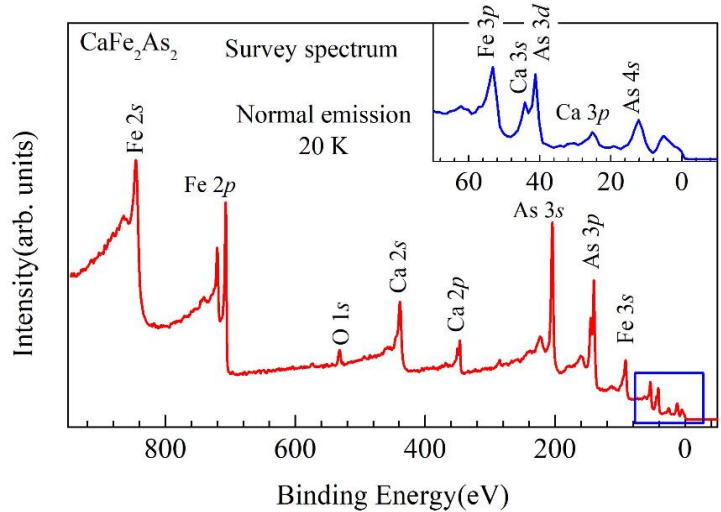

**FIGURE 2.** Survey spectrum collected at 20 K at normal emission geometry. Inset shows the boxed region in an enlarged scale. Except a small intensity due to O 1$s$ at about 531.2 eV due to adsorbed oxygens on the sample surface, all the feature in the spectrum correspond to the constituent elements.

## EXPERIMENT

High quality single crystals of CaFe$_2$As$_2$ was grown by high temperature solution growth method using Sn flux. The details of sample preparation is discussed in Ref. [6]. The quality of the crystals were characterized using powder $x$-ray diffraction, Laue diffraction and energy dispersive analysis of x-rays. HAXPES measurements were carried out at Petra III synchrotron radiation facility (Beamline - P09), DESY, Hamburg Germany. The photoelectron detection was done using Phoibos-150 electron analyzer. In order to get the best energy resolution, the experimental setup was optimized at different photon energies and the best resolution of 200 meV was achieved at a photon energy of 5947.5 eV. Samples were cleaved in ultrahigh vacuum condition and the measurements were done in the photoemission chamber at a pressure of 2×10$^{-10}$ torr. The position of the detector and photon source were fixed, and we rotated the sample to tune the depth sensitivity of technique from normal emission (bulk sensitive) to 60º angled emission (surface sensitive) [7]. The HAXPES measurements were done at three sample temperatures 210 K, 118 K and 20 K using an open cycle helium cryostat.

## RESULTS AND DISCUSSIONS

In Fig. 2, we show the survey spectrum collected at 20 K at normal emission geometry. For clarity, the features in the box is shown in an expanded energy scale in the inset. The Fermi level has been determined from the valence band spectrum of gold mounted in electrical contact with the sample in the same experimental setup. The binding energies of the features are determined using experimentally derived Fermi level as the reference energy and listed in Table I. All the features observed in the spectrum correspond to the photoemission signal from the energy levels of the constituent elements except a small impurity feature observed at 531.2 eV due to O 1$s$. Binding energy and intensity of the O 1$s$ signal indicates presence of a small amount of adsorbed oxygen at the surface and the bonding of the oxygen with the surface atoms is weak. The binding energies of all the measured core level features is compared with those found in other compounds and elemental metals. It appears the binding energy of Ca correspond to Ca$^{2+}$ charge state. However, the Fe and As atoms in the quasi 2D-layers possess core level binding energies very similar to the values observed in respective elemental metals [8].

In Fig. 3, we show the As 3$p$ core level spectra measured at both normal and 60º angled emission geometries at 210 K sample temperature. At this high photon energy, the escape depth of photoelectrons is close to 40 Å at normal emission. It becomes 20 Å at 60° emission angle. Thus, the surface contribution in measured spectrum will increase from about 14% to close to 26% with the change of emission angle from normal to 60°. Since, the As layers hybridize with the Ca atoms as also found in various other systems [9], any change in topmost surface layer is expected to have an impact in the properties of As layer in proximity of the surface layer in comparison to the properties of bulk As. Moreover, it is found that 122 class of materials cleave along the alkaline earth layer leaving 50% of the atoms on each of the cleaved surface [10]. The surface atoms reorganize to form linear structure. Thus, the As layer close to the cleaved surface is expected to have significant change.

**TABLE 1.** Binding Energy (BE) (in eV) of all core level state of Ca122 obtained from normal emission HAXPES measurement.

| Core Level | BE (eV) | Core Level | BE (eV) |
|---|---|---|---|
| Fe 2$s$ | 846.0 | Fe 3$s$ | 91.6 |
| Fe 2$p_{1/2}$ | 719.8 | Fe 3$p_{1/2}$ | 54.6 |
| Fe 2$p_{3/2}$ | 706.8 | Fe 3$p_{3/2}$ | 53.0 |
| Ca 2$s$ | 438.8 | Ca 3$s$ | 44.0 |
| Ca 2$p_{1/2}$ | 350.1 | As 3$d_{3/2}$ | 41.7 |
| Ca 2$p_{3/2}$ | 346.4 | As 3$d_{5/2}$ | 41.0 |
| As 3$s$ | 204.5 | Ca 3$p$ | 24.8 |
| As 3$p_{1/2}$ | 145.4 | As 4$s$ | 12.0 |
| As 3$p_{3/2}$ | 140.4 | | |

In our experiments, with the change in surface sensitivity (about 12% increase), we do not observe noticeable change in the measured spectra. This indicates that surface-bulk difference in the properties of As is either absent or not large enough to be detected by such small changes in surface sensitivity. The simulation of the measured normal emission As $3p$ core level spectrum using asymmetric Gaussian-Lorentzian product function is shown in Fig. 3 (lower panel). From the intensity ratio, peak energy and the energy differences between different peaks, it is clear that the sharp and intense features are As $3p$ spin-orbit split peaks with the splitting of about 5.0 eV. The broad features around the binding energies of 160 eV and 175 eV can be simulated by two sets of peaks with similar energy separation and intensity ratio of As $3p$ spin-orbit split peaks. Such features at similar energy separation from the main photoemission signal are also observed in other core level spectra as evident in the survey spectrum shown in Fig. 2. Thus, these features can be attributed to the energy loss due to collective excitations such as plasmons [7].

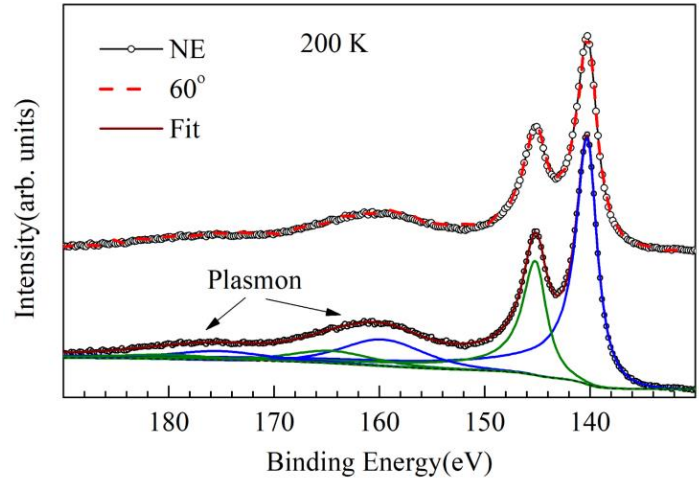

FIGURE 3. As $3p$ spectra taken at normal and $60°$ angled emission geometries at 210 K. Theoretical simulation of measured normal emission spectrum.

The simulation of the experimental data shows finite asymmetry in the lineshape of the core level peaks. Such asymmetry often arises due to excitation of valence electrons across the Fermi level along with the core level excitations. The presence of large asymmetry indicates that the contribution of As $4p$ partial density of states at the Fermi level is finite, which appears due to hybridization with the Fe $3d$ valence states [9]. This observation emphasizes the important role of ligand states in the electronic properties of this system as found in Fe-based systems [11].

In Fig. 4, we show As $3p$ spectra collected at 25 K, 118 K and 210 K sample temperatures for both the emission angles. This wide temperature range spans across the magnetic and structural transition temperature of 170 K. In the orthorhombic phase, Fe-As bond length reduces from 2.410 Å to 2.388 Å and the As-Ca bond length increases from 3.052 Å at 250 K to 3.176 Å at 50 K [1]. Thus, it is expected that the hybridization of As with Fe and Ca will change significantly. However, we do not observe any change in the experimental spectra in the whole temperature range studied in both normal emission and 60° angled emission cases. It is to be noted here that the formation of spin-density wave state below 170 K opens up an energy gap in one of the three bands forming hole pocket around $\Gamma$ point [6] and signature of nematicity is observed at low temperatures [12]. Therefore, absence of any change in asymmetry and/or peak position of the As 3*p* core level spectra is curious. Evidently more studies are required to understand this puzzling scenario.

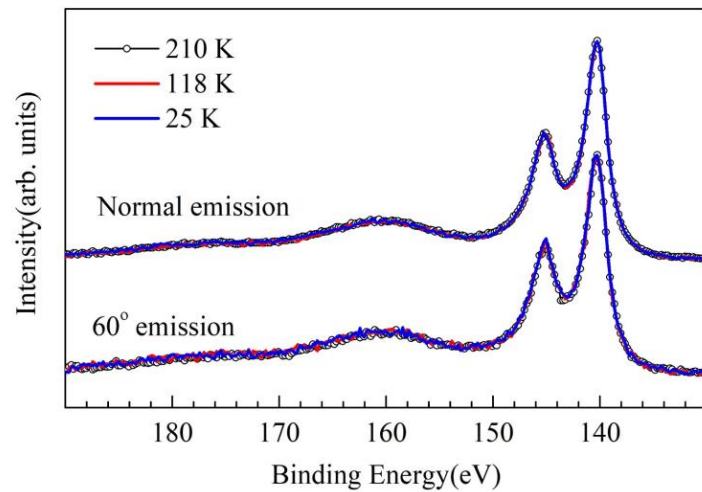

FIGURE 4. As $3p$ normal and $60°$ angled HAXPES spectra taken at 210 K, 118 K and 25 K sample temperatures.

## CONCLUSIONS

In summary, we studied the role of ligand states in the electronic properties of CaFe$_2$As$_2$ employing hard *x*-ray photo-electron spectroscopy at two different photoelectron emission angles and varied sample temperatures. No surface-bulk difference is observed in the As 3*p* core level spectra studied with the change of emission angle even though the As layer close to the surface Ca layer is expected to be influenced by 50% populated cleaved surface. Asymmetric line shape of As 3*p* core level spectra indicates finite As 4*p* - PDOS in the vicinity of Fermi level which has been observed from theoretical calculations due to Fe-As hybridization. Two sets of plasmon peaks at binding 160.1 eV and 175.9 eV are observed. No spectral change could be observed in our measurements with change in temperature from 210 K to 25 K although a structural and magnetic phase transition occurs at 170 K.

## ACKNOWLEDGMENTS

Authors acknowledge the financial support from DST-DESY and thank Dr. Andrei Gloskovskii for his support during the experiments. KM acknowledges financial assistance from the Department of Science and Technology, government of India under J. C. Bose Fellowship program and the Department of Atomic Energy under the DAE-SRC-OI Award program.